\documentclass{lncse}
\usepackage[]{graphicx}
\usepackage{amssymb}
\usepackage{latexsym}

\begin{document}

\title{A new cellular automata model for city traffic}

\author{A. Schadschneider \inst{1} \and D. Chowdhury \inst{1,2} \and
  E. Brockfeld \inst{3} \and K. Klauck \inst{1} \and L. Santen \inst{1}
\and J. Zittartz \inst{1}}

\institute{Institut f\"ur Theoretische Physik, Universit\"at zu 
K\"oln, 50937 K\"oln, Germany 
\and
Physics Department, I.I.T., Kanpur 208016, India
\and
Institut f\"ur Umweltsystemforschung,
Universit\"at Osnabr\"uck, 49076 Osnabr\"uck, Germany}

\maketitle

\begin{abstract}
We present a new cellular automata model of vehicular traffic 
in cities by combining ideas borrowed from the Biham-Middleton-Levine 
(BML) model of city traffic and the Nagel-Schreckenberg (NaSch) model 
of highway traffic. The model exhibits a dynamical phase transition 
to a completely jammed phase at a critical density which
depends on the time periods of the synchronized signals.
\end{abstract}

\section{Introduction}

A one-dimensional cellular automata (CA) model of highway traffic and 
a two-dimensional CA model of city traffic were developed 
independently by Nagel and Schreckenberg (NaSch) \cite{NaSch} and Biham, 
Middleton and Levine (BML) \cite{BML}, respectively\footnote{For a 
review of the different approaches in modelling traffic flow we refer 
to \cite{review} and references therein.}.
Highway traffic becomes gradually more and more congested in the 
NaSch model with the increase of density. Traffic jams appear because 
of the {\it intrinsic stochasticity} of the dynamics but no  
jam persists for ever.
On the other hand, a first order phase 
transition takes place in the BML model at a finite non-vanishing 
density, where the average velocity of the vehicles vanishes 
discontinuously signalling complete jamming. In the BML model, the 
randomness arises only from the 
{\it random initial conditions}, as the dynamical rule for the 
movement of the vehicles is fully deterministic \cite{BML}. 

In the BML model a square lattice models the network of the streets.
Each site of the lattice represents a crossing of an east-bound 
and a north-bound street; it can be empty or occupied by a vehicle 
moving either to the east or to the north. The dynamics is controlled 
by signals of period 1 such that at odd time steps only the 
east-bound vehicles are updated whereas at even time steps only 
the north-bound vehicles are updated, both in parallel. 
During the updating a vehicle moves one site forward
if the next site ahead is not occupied by any other vehicle.
In this simplest version of the model lane changes (e.g.\ by turning)
are not possible and the number of vehicles on each street is conserved 
separately.

In the NaSch model a highway is represented by a one-dimensional 
lattice of cells that can accomodate not more than one vehicle at a time.
Each vehicle is characterized by a maximum velocity
$v_{max}$ and a randomization parameter $p$. The dynamics consists
of four steps, each applied in parallel to all vehicles. In the
first step all vehicles accelerate by 1 if they have not already reached
the maximum velocity $v_{max}$. Step 2 is the interaction step.
Vehicles which have $d$ empty cells in front and a velocity $v>d$ reduce
their velocity  to $v=d$ in order to avoid a crash. In step 3 the
velocity is reduced by one unit with probability $p$. In step 4 the
vehicles move forward $v$ cells where $v$ is the new velocity after 
the randomization step 3.

\section{Definition of the model}

\begin{figure}[ht]
  \begin{center}
    \includegraphics[width=.5\textwidth]{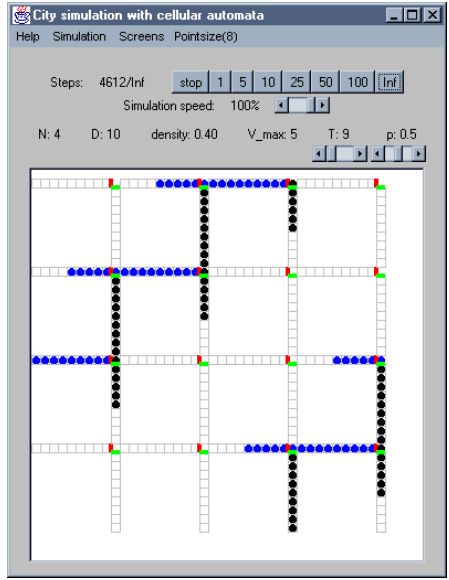}
    \caption{Structure of the underlying lattice for $N=4$ and $D=10$.
Shown is a typical jammed configuration of the vehicles.
The east-bound and north-bound vehicles are represented by the
blue and black symbols, respectively.}
    \label{fig:gridlock}
    \end{center}
\end{figure}

In the BML model the interplay between the vehicle dynamics and
the time-scale set by the length of the signal period can not be
studied. We therefore suggested a "unified" model \cite{cs} combining 
the BML model with the rules for the vehicle dynamics of the NaSch model.
The lattice of our new model consists (in the simplest case) of
$N$ north-bound and $N$ east-bound streets. The $N\times N$ crossings
of these streets are arranged equidistantly. Between two consecutive
crossings on a street there are $D-1$ cells, i.e.\ each street has
length $L = ND$ (see Fig.\ \ref{fig:gridlock}). 
The signals, installed at the crossings, are synchronized in such a way 
that all the signals remain green for the east-bound vehicles (and 
simultaneously, red for the north-bound vehicles) for a time interval 
$T$ and then, simultaneously, all the signals turn red for the east-bound 
vehicles (and green for the north-bound vehicles)
for the next $T$ time steps before turning green again. 
This process is repeated so that there is a total time 
interval $2T$ between the beginning of two successive 
green (or red) phases of the signals. 

As in the NaSch model the speed $v$ of each vehicle can take one 
of the $v_{max}+1$ integer values $v=0,1,...,v_{max}$.
Suppose, $v_n$ is the speed of the $n$-th vehicle at time $t$ 
while moving either towards east or towards north. 
In the initial state of the system, $N_x$ ($N_y$) vehicles are 
distributed among the east-bound (north-bound) streets. Here
we only consider the case $N_x = N_y =N_v/2$ where $N_v$ is the
total number of vehicles. Since in the initial configuration the
occupation of a crossing is strictly avoided, the global density
is defined by $\rho=N_v/N^2(2D-1)$.
Also, suppose $d_n$ is the distance to the next vehicle in front
while $s_n$ denotes the distance to the nearest crossing 
in front of it.
At each {\em discrete time} step $t \rightarrow t+1$, the arrangement of 
vehicles is updated {\em in parallel} according to the following 
"rules":
\begin{itemize}
\item Step 1 (Acceleration):\\
$v_n \rightarrow \min(v_n+1,v_{max})$\\
\item Step 2 (Deceleration due to other vehicles or signals):\\
\begin{description}
\item{Case I:} The signal is {\bf red} for the $n$-th vehicle under
consideration:\\
$v_n \rightarrow \min(v_n,d_n-1,s_n-1)$
\item{Case II:} The signal is {\bf green} for the $n$-th vehicle under
consideration:\\
If the signal is going to turn to red in the next timestep then\\             
\phantom{else }$v_n \rightarrow \min ( v_n,d_n-1,s_n -1)$\\
else $v_n \rightarrow \min ( v_n, d_n-1)$.\\
\end{description} 
\item Step 3 (Randomization):\\
$v_n \rightarrow \max(v_n - 1,0)$ with probability $p$\\
\item Step 4 (Movement):\\
$x_n \rightarrow x_n+v_n$.
\end{itemize}

Note that we have simplified Case II of Step 2 in comparison to \cite{cs}.
This simplification does not change the overall behaviour of the 
model \cite{csprep}.

These rules are not merely a combination of the BML and the NaSch 
rules but also involve some modifications.
For example, unlike all the earlier BML-type models, a vehicle 
approaching a crossing can keep moving, even when the signal is 
red, until it reaches a site immediately in front of which there 
is either a halting vehicle or a crossing. Moreover, if $p=0$ 
every east-bound (north-bound) vehicle can adjust speed in the 
deceleration stage so as not to block the north-bound (east-bound) 
traffic when the signal is red for the east-bound (north-bound) 
vehicles. 


\section{Results}

\begin{figure}[hb]
\noindent
\begin{minipage}[b]{0.99\linewidth}
    \centering
    \includegraphics[width=0.6\linewidth]{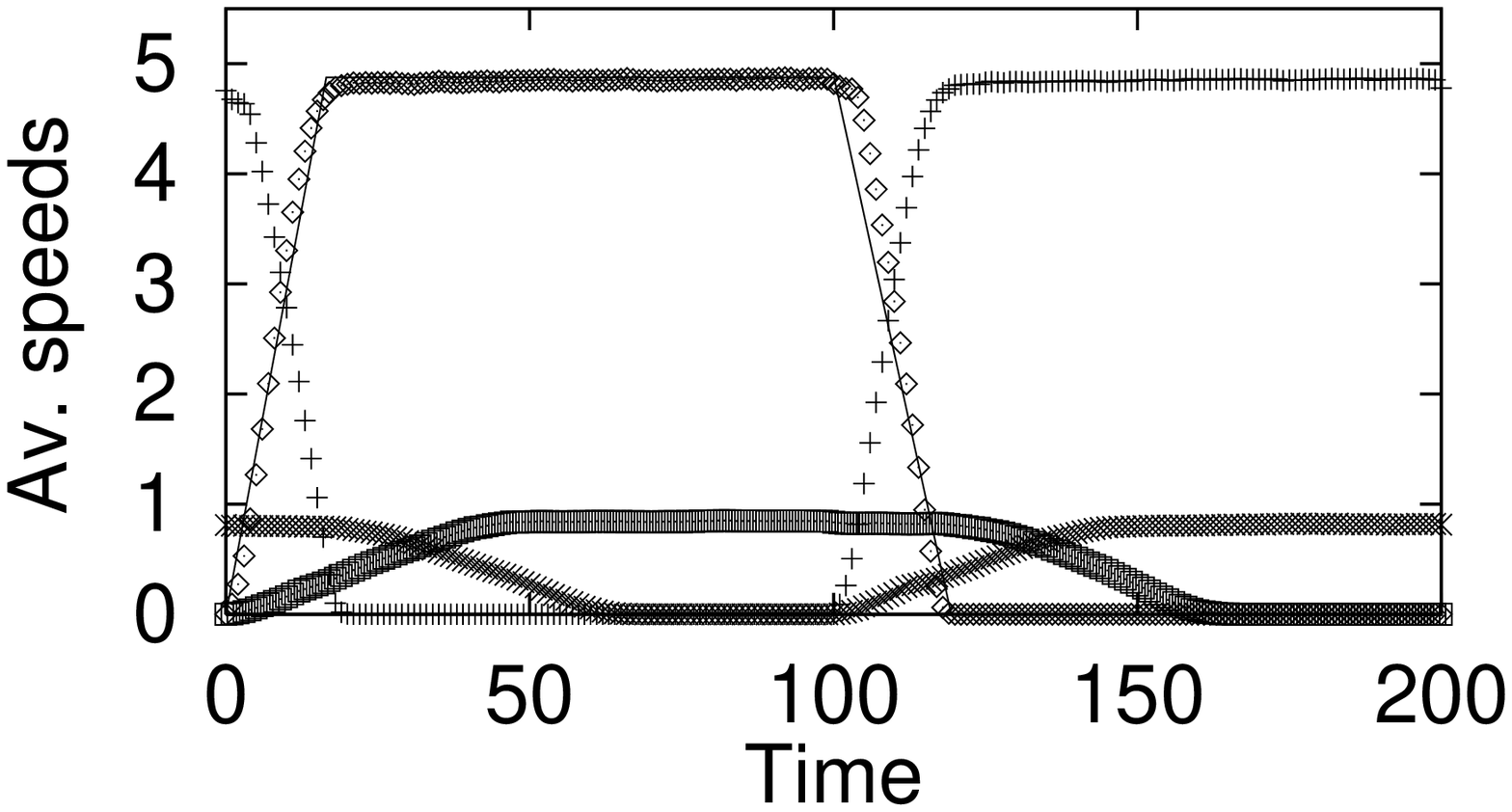}
    \caption{Time-dependence of average speeds of vehicles. The symbols
$+$, $\times$, $\ast$ and $\Box$ correspond, respectively, to the
average speeds $\langle v_x\rangle$, $\langle v_y\rangle$, and the 
fractions of vehicles with instantaneous speed $V = 0$, $f_{x0}$ 
and $f_{y0}$, respectively. 
The common parameters are $v_{max} = 5, p = 0.1, D = 100$,
$T = 100$ and $c = 0.1$. The continuous line has been obtained from 
heuristic arguments given in [4].}
    \label{fig:avveloc}
\end{minipage}
\centering
    \includegraphics[width=0.7\linewidth]{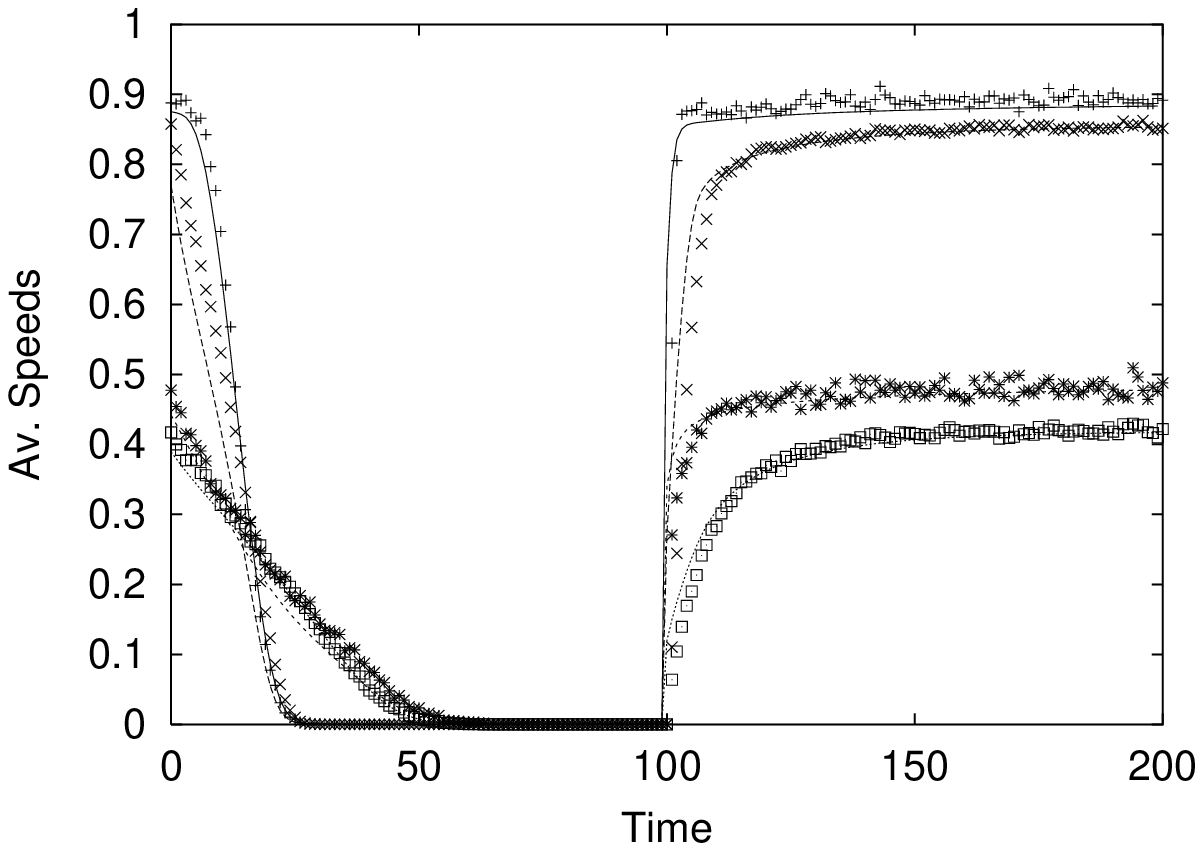}
    \caption{Comparison between MC data and 2-cluster results. The common
      paramters are $v_{max}=1,$ $D=25$, $T=100$ and $N=4$. The solid
      lines correspond to 2-cluster results. The symbols
      $+,\times,\ast,\boxdot$ correspond to the $\rho/p$ MC data sets
      $0.05/0.1,0.25/0.1,0.05/0.5,0.25/0.5$ respectively.}
\label{fig:cluster}
\end{figure}

The variations of $\langle v_x \rangle$ and $\langle v_y
\rangle$ with time (see Fig.~\ref{fig:avveloc}) 
as well as with $c$, $D$, $T$ and $p$ in the flowing phase are 
certainly more realistic than in the BML model \cite{cs}.
In the case of $\rho<\rho_c$ and for $v_{max}=1$
the dynamics of the system can be described accurately by treating
a single street with an improved version of the 2-cluster
approximation \cite{tgf},
where the 2-cluster probabilities  are equipped with a time and space
dependence \cite{csprep} (see Fig.~\ref{fig:cluster}) . 

\begin{figure}[ht]
  \begin{center}
    \includegraphics[width=.65\textwidth]{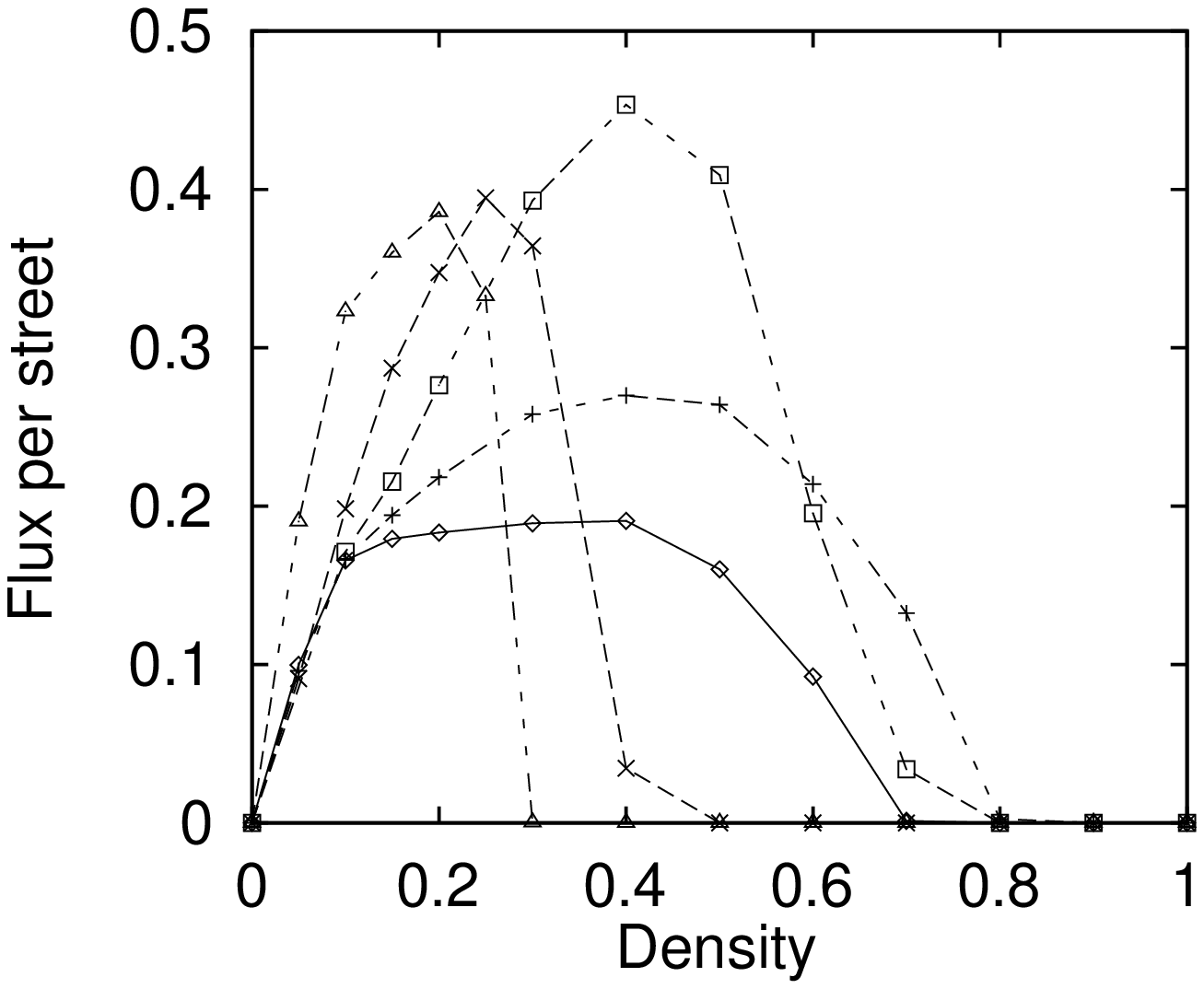}
    \caption{Fundamental diagram for $v_{max} = 5, p = 0.5$, $L=100$, 
and $D = 20$. The symbols $\diamond, +, \square, \times$ and $\triangle$ 
correspond, respectively, to $T = 100, 50, 20, 10, 4$.}
    \label{fig:fund}
    \end{center}
\end{figure}

The fundamental diagram (Fig.~\ref{fig:fund}) also shows a rather complex
behaviour, at least for finite systems. E.g.\ the density corresponding 
to the maximum flux shifts to smaller densities 
with the decrease of $T$. Furthermore, the maximum throughput 
is a non-monotonic function of $T$ in the "free-flowing" phase; 
this result may be of practical use in traffic engineering 
for maximizing the throughput. 

A phase transition from the "free-flowing" dynamical phase to the
completely "jammed" phase takes place in this model at a vehicle
density $\rho_c$. The intrinsic
stochasticity of the dynamics, which triggers the onset of jamming, is
similar to that in the NaSch model, while the phenomenon of complete
jamming through self-organization as well as the final jammed
configurations (see Fig.~\ref{fig:gridlock}) are similar to those in 
the BML model.  

\begin{figure}[ht]
  \begin{center}
    \includegraphics[width=.5\textwidth]{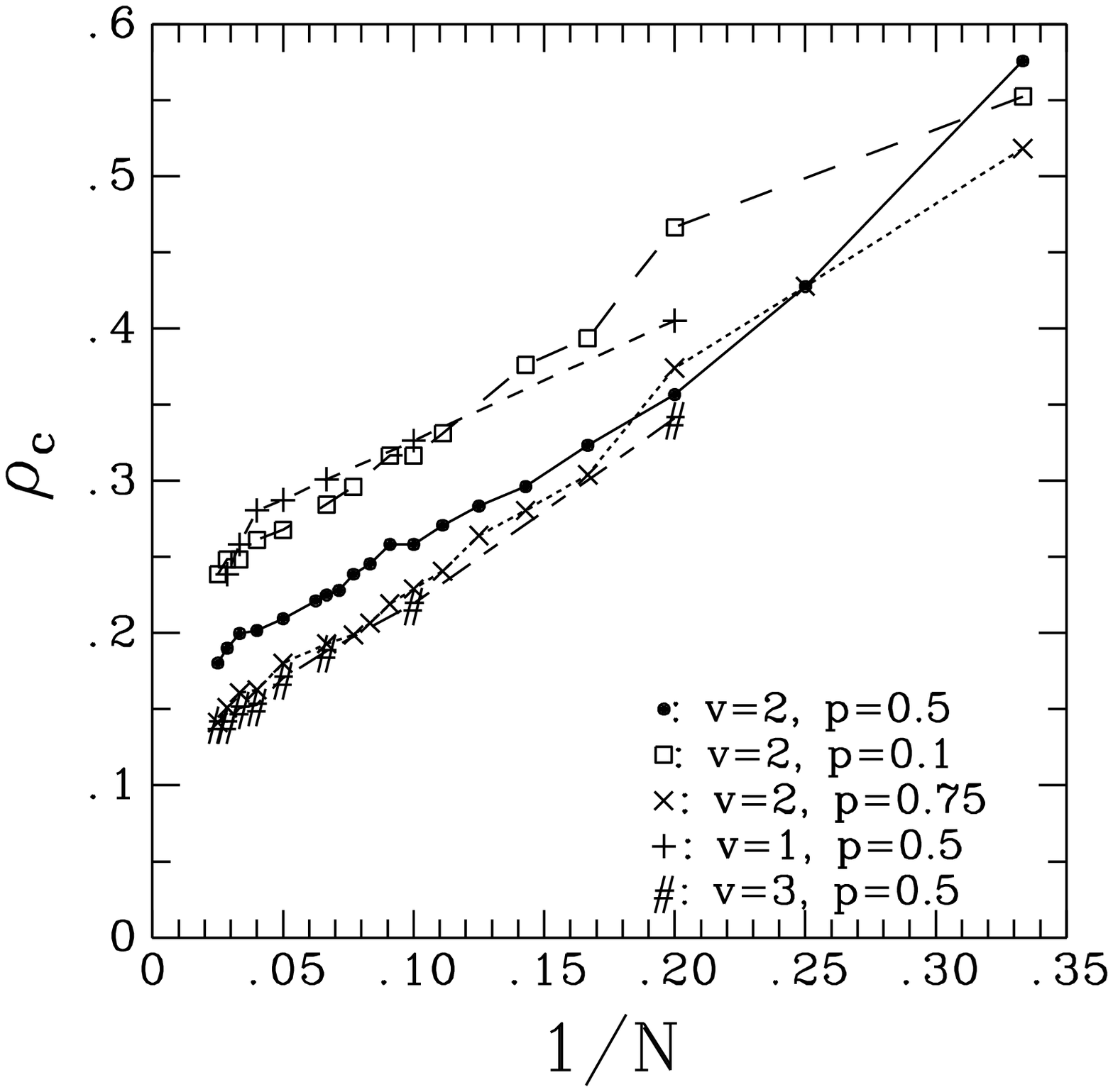}
    \caption{Critical density $\rho_c$ for different parameter
combinations
     as function of the number of streets $N$. The common parameters are
     $D=20$ and $T=5$.}
    \label{fig:rhoc}
    \end{center}
\end{figure}

Due to the importance of finite-size and finite-time corrections it is
not clear up to now how the critical density $\rho_c(D)$ depends on 
the dynamical parameters  $v_{max}$, $p$ and $T$.
It is possible that in the thermodynamic limit $N\to\infty$ the density
$\rho_c$ is completely determined by the structure
of the underlying lattice, i.e.\ by $D$, as long as $p>0$ which is
necessary for the jamming transition to occur. In that case $\rho_c$
would be independent of $v_{max}$, $p$ and $T$ and the transition
would be of 'geometrical' nature similar to the percolation transition.
On the other hand, the transition could also be truly dynamical with
$\rho_c$ depending also on $v_{max}$, $p$ or $T$.
The data obtained so far from the computer simulations (see 
Fig.~\ref{fig:rhoc}) do not conclusively rule out either of these 
two possible scenarios.

The "unified" model has been formulated intentionally to keep 
it as simple as possible and at the same time capture some of the
interesting features of the NaSch model as well as the BML model. 
We believe that this model can be generalized (i) to allow 
traffic flow in both ways on each street which may consist of 
more than one lane, (ii) to make more realistic rules for the 
right-of-the-way at the crossings and turning of the vehicles, 
(iii) to implement different types of synchronization or 
staggering of traffic lights, e.g.\ green-waves \cite{simon}. 

\section*{Acknowledgment}

Part of this work has been supported by the SFB 341 
(K\"oln-Aachen-J\"ulich).


\begin{thebibliography}{99}

\bibitem{NaSch} K.\ Nagel and M. Schreckenberg: 
J. Physique I, {\bf2}, 2221 (1992)

\bibitem{BML} O. Biham, A.A. Middleton and D. Levine: 
Phys. Rev. A {\bf 46}, R6124 (1992)

\bibitem{review} D. Chowdhury, L.\ Santen and A. Schadschneider:
to be published in Phys.\ Rep.



\bibitem{cs} D. Chowdhury and A. Schadschneider: 
Phys. Rev. E {\bf 59}, R1311 (1999) 

\bibitem{csprep} 
D. Chowdhury, K. Klauck, L. Santen, A. Schadschneider and 
J. Zittartz: to be published

\bibitem{tgf} A. Schadschneider, in 
  M.\ Schreckenberg and D.E.\ Wolf (Eds.): {\em Traffic and Granular
    Flow '97}. Springer, Singapore (1998)

\bibitem{simon} P.M.\ Simon and K.\ Nagel:
Phys.\ Rev.\ E {\bf 58}, 1286 (1998)
  

\end{thebibliography}
\end{document}